\documentclass[aps,prb,twocolumn,amssymb,amsmath,showpacs,superscriptaddress]{revtex4-1}

\usepackage{bm}
\usepackage{times}
\usepackage{graphicx}
\usepackage{color}
\usepackage{dcolumn}
\usepackage[colorlinks=true, letterpaper=true, pdfstartview=FitV, linkcolor=blue, citecolor=blue, urlcolor=blue]{hyperref}
\usepackage{appendix}
\usepackage[normalem]{ulem}

\begin{document}
\title{Unconventional Hall Effect in Gapless Superconductors: Transverse Supercurrent Converted from Normal Current}

\author{Miaomiao Wei}
\thanks{These authors contributed equally to this work.}
\affiliation{College of Physics and Optoelectronic Engineering, Shenzhen University, Shenzhen 518060, China}

\author{Longjun Xiang}
\thanks{These authors contributed equally to this work.}
\affiliation{College of Physics and Optoelectronic Engineering, Shenzhen University, Shenzhen 518060, China}

\author{Fuming Xu}
\affiliation{College of Physics and Optoelectronic Engineering, Shenzhen University, Shenzhen 518060, China}
\affiliation{Quantum Science Center of Guangdong-Hongkong-Macao Greater Bay Area (Guangdong), Shenzhen 518045, China}

\author{Bin Wang}
\email[]{wangbin@szu.edu.cn}
\affiliation{College of Physics and Optoelectronic Engineering, Shenzhen University, Shenzhen 518060, China}

\author{Jian Wang}
\email[]{jianwang@hku.hk}
\affiliation{College of Physics and Optoelectronic Engineering, Shenzhen University, Shenzhen 518060, China}
\affiliation{Quantum Science Center of Guangdong-Hongkong-Macao Greater Bay Area (Guangdong), Shenzhen 518045, China}
\affiliation{Department of Physics, The University of Hong Kong, Pokfulam Road, Hong Kong, China}

\begin{abstract}
A normal metallic system proximitized by a superconductor can exhibit a gapless superconducting state characterized by segmented Fermi surfaces, as confirmed experimentally. In such a state, quasiparticle states remain gapless along one direction, while a superconducting gap opens in the perpendicular direction. This anisotropy enables a novel Hall effect in gapless superconductors, termed the superconducting Hall effect (ScHE), where a longitudinal normal current carried by quasiparticles is converted into a dissipationless transverse supercurrent. Employing both the thermodynamic approach for bulk systems and quantum transport theory for a four-probe setup, we demonstrate the existence of this effect and reveal its intrinsic origin as the quasiparticle Berry curvature. The predicted ScHE can be experimentally verified via the standard angular-dependent Hall measurements performed on gapless superconductors.
\end{abstract}

\maketitle

\noindent{\it Introduction} --- Hall effects, characterized by transverse responses to longitudinal driving forces, represent a major research branch in condensed matter physics. Hall transport measurements provide powerful diagnostic tools for probing material properties and exploring phases of matter. The occurrence of Hall effects is generically tied to symmetry breaking. For instance, both quantum Hall effect\cite{QHE} and anomalous Hall effect\cite{AHE1,AHE3,Nagaosa-AHE} require the breaking of time-reversal symmetry ($\cal{T}$), which are intrinsic effects arising from the geometric properties of Bloch bands, namely Berry phase and Berry curvature,\cite{D-Xiao} respectively. In systems preserving $\mathcal{T}$ but breaking inversion symmetry ($\mathcal{P}$), an extrinsic second-order nonlinear Hall effect induced by Berry curvature dipole can emerge.\cite{Sodemann2015,Guinea2015,SXu2018,QMa2019,KKang2019,Kumar2021,Du2021} In contrast, in systems that break both $\mathcal{P}$ and $\mathcal{T}$ while preserving combined $\mathcal{P}\mathcal{T}$ symmetry, such as certain antiferromagnets, the intrinsic second-order Hall effect has been identified,\cite{Wang2023,Trevisan2023} which is rooted in the quantum metric dipole.\cite{Gao2014,Wang2021,Liu2021,Yan2024} These discoveries highlights the critical role of band geometry in transport phenomena and enriches our understanding on topological phases of quantum materials.


Superconductivity is a remarkable macroscopic quantum state, and its interplay with Hall physics has been an intriguing topic of research. Recently, a novel superconducting state---known as the gapless superconductor---has been proposed and experimentally realized.~\cite{L-Fu1,L-Fu2, L-Fu3,J-Jia} In this phase, the Fermi surface of Bogoliubov quasiparticles partially intersects the superconducting gap in momentum space, resulting in the coexistence of gapless quasiparticle states and gaped Cooper pair condensate at the same energy. This segmented Fermi surface allows quasiparticles to propagate along one direction while Cooper pairs flow in the orthogonal direction, without metal-superconductor phase transition. The characteristic band structure of gapless superconductors naturally raises the question of whether a longitudinal quasiparticle current can be converted into a transverse supercurrent through a Berry curvature-like mechanism. In this work, we demonstrate that such an unconventional effect can indeed occur in gapless superconductors. This phenomenon, referred to as the superconducting Hall effect (ScHE), is an intrinsic effect induced by the Berry curvature of Bogoliubov quasiparticles.

The underlying mechanism of this ScHE is fundamentally distinct from previously studied Hall effects in superconductors. Several mechanisms have been reported in literatures.\cite{Parafilo2023,Sonowal2024,Iwasa2022,Daido2024,Mironov2024,Dong2025} For instance, in photoinduced Hall effects predicted in two-dimensional superconductors,\cite{Parafilo2023,Sonowal2024} Cooper pairs of the built-in supercurrent are excited by incident light, leading to either a transverse supercurrent response within the superconducting gap\cite{Parafilo2023} or nonlinear ac quasiparticle Hall current out of the gap.\cite{Sonowal2024} Near the metal-superconductor transition criticality, the second-order nonlinear Hall responses can be significantly enhanced by superconducting fluctuations,\cite{Iwasa2022,Daido2024} where fluctuating Cooper pairs with a divergent relaxation time dominate quasiparticle transport.\cite{Dong2025} Additionally, an electromagnetic wave incident on a superconductor surface can induce a dc supercurrent and second-harmonic response due to the photon drag effect.\cite{Mironov2024} In contrast, the ScHE proposed in this work originates intrinsically from the band geometry of gapless superconductors and does not involve the metal-superconductor phase transition; it can be driven by a small electric field in periodic bulk systems and requires no optical excitation. Our findings predict an unconventional supercurrent Hall effect, which can be experimentally verified through angular-dependent Hall measurements.



\begin{figure}
\centering
\includegraphics[width=\columnwidth]{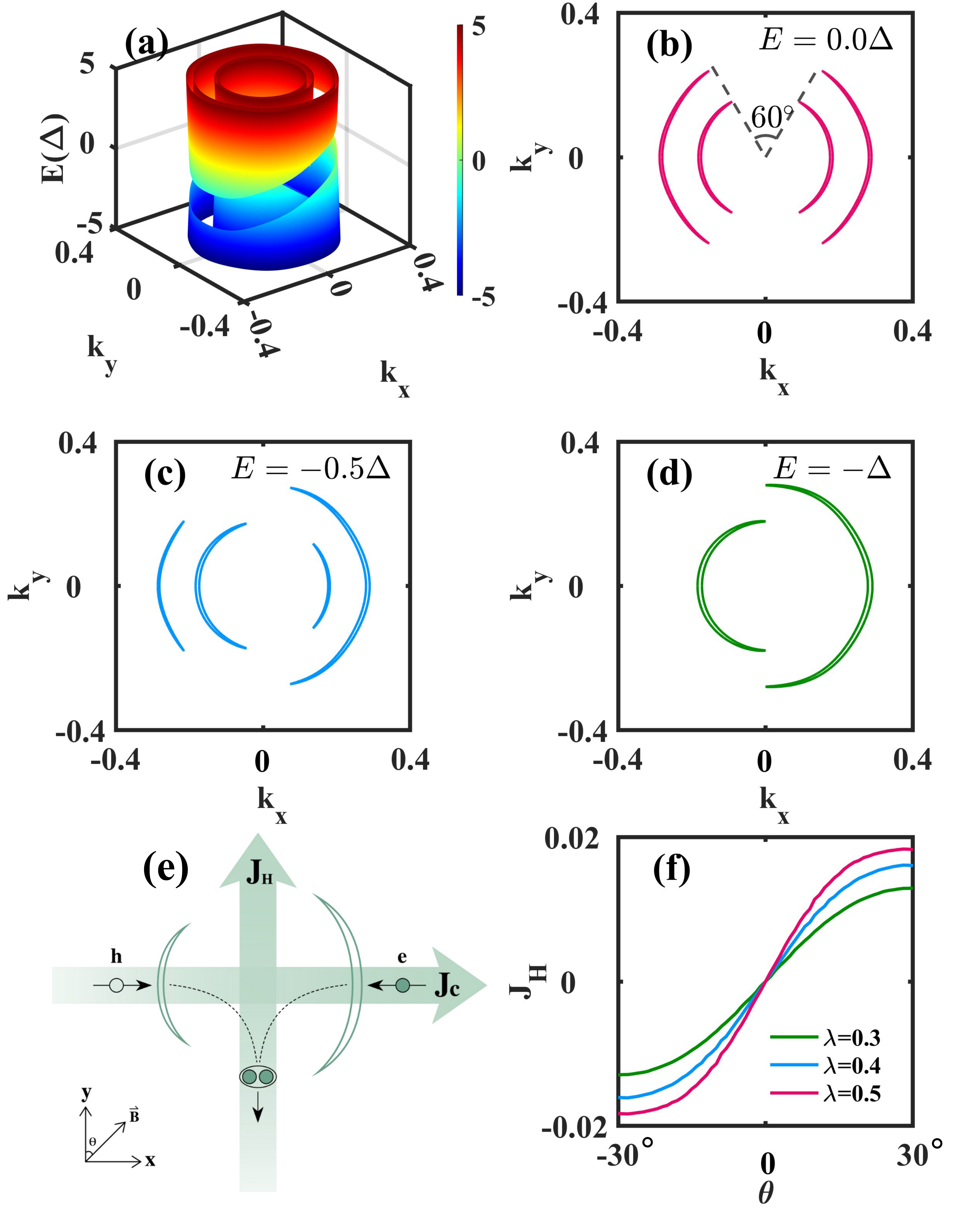}
\caption{(a) Quasiparticle band structure of the gapless superconducting state for the BdG Hamiltonian in Eq.~(\ref{ham}). (b-d) The segmented Fermi surfaces for $E=0$, $-0.5\Delta$, $-\Delta$. (e) Schematic of the superconducting Hall effect. An electric field drives a longitudinal quasiparticle current $J_c$ in the $x$ direction; meanwhile, the motion of electrons and holes is deflected due to the Berry curvature effect and they are combined into finite-momentum Cooper pairs, resulting in a transverse supercurrent along the $y$ direction. (f) Superconducting Hall current as a function of the orientation angle $\theta$ of the in-plane magnetic field for different warping strength $\lambda$, which is calculated with Eq.~(\ref{current3}). Other parameters are $t=1$, $\mu=0.05t$, $\alpha_R=0.1t$, $\lambda=0.5t$, $\Delta=0.001t$, $B_x =0$, $B_y = 2\Delta$.\cite{L-Fu1} } \label{FIG1}
\end{figure}

\smallskip
\noindent{\it Mechanism of the ScHE} --- We begin by theoretically analyzing the emergence of the ScHE based on the the gapless superconducting state in a two-dimensional (2D) superconductor. Our analysis starts with the following single-particle Hamiltonian\cite{S-Sun,J-Jia}
\begin{eqnarray}
H_0 &=& \sum_{k}\Big[ {{\xi}_k}+{{\alpha}_R}( {k_x}{{\sigma}_{y}}-{k_y}{{\sigma}_x} )+\emph{g}\mu_B (B_x \sigma_x + B_y \sigma_y) \nonumber \\
&+& \lambda \left( k_{x}^{3}-3{k_x}k_{y}^{2} \right){{\sigma}_{z}} \Big],\label{ham0}
\end{eqnarray}
where ${\xi}_k={\hbar^{2}}{{{k}^{2}}}/{2m} -\mu$ is the kinetic energy measured from the chemical potential $\mu$. The second term represents Rashba spin-orbit coupling (SOC) with strength ${\alpha}_{R}$ and Pauli matrix $\sigma_i$ ($i=x,y,z$) for spin. The third term describes the Zeeman effect induced by in-plane magnetic filed $B_{x/y}$, where $\emph{g}$ is the effective $\emph{g}$ factor and $\mu_B$ is the Bohr magneton. The magnetic field is oriented at an angle $\theta$ with respect to the $y$ axis, as shown in Fig.~\ref{FIG1}(e). The last term is the hexagonal warping contribution with coefficient $\lambda$,\cite{HWE1,HWE2,HWE3} which enables nonzero Berry curvature in this Hamiltonian.

When the system is proximity-coupled to an $s$-wave superconductor, the corresponding Bogoliubov-de Gennes (BdG) Hamiltonian in the Nambu basis $( {c}_{\textbf{k},\uparrow},{c}_{\textbf{k},\downarrow},c_{-\textbf{k},\uparrow}^{\dagger},c_{-\textbf{k},\downarrow}^{\dagger})$ takes the form\cite{SC-Zhang}
\begin{equation}
H_{BdG} = \left( \begin{matrix}
   {H_0}( \textbf{k} ) & i\Delta{{\sigma}_y}  \\
   -i\Delta{{\sigma}_y} & -H_{0}^{*}( -\textbf{k} )  \\
\end{matrix} \right), \label{ham}
\end{equation}
where $\Delta$ is the proximity-induced superconducting gap. This Hamiltonian can support a gapless superconducting phase,\cite{L-Fu1,L-Fu2,L-Fu3,M-Wei1} which has been experimentally realized in Bi$_2$Te$_3$ thin films proximitized by superconductor NbSe$_2$.\cite{J-Jia}


The quasiparticle band structure for the BdG Hamiltonian Eq.~(\ref{ham}) is illustrated in Fig.~\ref{FIG1}(a). The presence of finite SOC and an in-plane magnetic field breaks the $\mathcal{PT}$ symmetry, leading to the asymmetric energy dispersion $E(\mathbf{k}) \neq E(-\mathbf{k})$ and a tilted superconducting gap. Consequently, when the Fermi energy intersects this superconducting gap, a segmented Fermi surface will emerges. The evolution of these segmented Fermi surfaces with varying energy is depicted in Fig.~\ref{FIG1}(b)-(d). At $E = 0$, four ``banana-shaped'' Fermi pockets appear [Fig.~\ref{FIG1}(b)], where quasiparticle states exist only along the contours of these pockets. As the energy decreases to $E=-0.5\Delta$, two of the Fermi segments expand while the other two shrink, as shown in Fig.~\ref{FIG1}(c). At $E = -\Delta$, only two Fermi segments persist, and the superconducting gap is closed [Fig.~\ref{FIG1}(d)]. Upon further lowering the energy, two ends of the segments approach one another and eventually merge.\cite{supple} In the calculations of Fig.~\ref{FIG1}(a)-(d), the magnetic field is fixed along the $y$ direction. When the magnetic field is rotated, the orientation of the segmented Fermi surfaces rotates accordingly. The influence of the rotating magnetic field and the warping strength on the morphology of segmented Fermi surfaces is discussed in detail in the Supplementary Material.\cite{supple}

For the configurations shown in Fig.~\ref{FIG1}(b) and (c), quasiparticle transport is primarily permitted along the $x$ direction, while only supercurrent is allowed along the $y$ direction within the superconducting gap. This scenario satisfies the conditions for realizing the ScHE. The physical mechanism of ScHE is schematically illustrated in Fig.~\ref{FIG1}(e). When an electric field $E_x$ is applied, quasiparticles are driven out of the segmented Fermi pockets, generating a longitudinal quasiparticle current $J_c$. Meanwhile, the motion of quasiparticles is deflected due to the anomalous velocity arising from Berry curvature, as detailed described below; electrons and holes are combined into finite-momentum Cooper pairs,\cite{L-Fu3} inducing a transverse superconducting Hall current $J_H$. This ScHE represents an unconventional manifestation of the Hall effect, where longitudinal quasiparticle currents are converted into transverse supercurrents. It is worth noticing that the ScHE proposed here originates from the intrinsic band geometry of the gapless superconducting state and does not involve any phase transition. In the following, we further elucidate the transport characteristics of this ScHE.



\smallskip
\noindent{\it Theoretical framework for describing ScHE} --- Quasiparticle dynamics in superconductors has been studied with various theoretical frameworks. For example, the thermo-electric and thermo-spintronic responses of quasiparticles have been discussed with the semiclassical theory.\cite{quasiparticleBC1,quasiparticleBC2} The optical responses of parity/inversion-breaking superconductors under light irradiation have been investigated using the response theory.~\cite{JEMooreSC,YanaseSC} In this work, we focus on the transport properties of gapless superconducting states driven by an external electric field. We develop a unified theoretical framework capable of describing both quasiparticle current and supercurrent, which coexist in gapless superconductors. The conventional definition of current density, $J = \sum_n \int v_n g_n dk$ where $v_n$ is the velocity and $g_n$ is the nonequilibrium distribution function of the $n$-th band, is inadequate for describing the supercurrent, as the velocity of Cooper pairs is not well defined in the BdG Hamiltonian. Meanwhile, it is essential to treat the contributions from quasiparticles and Cooper pairs on equal footing.

We notice that the gapless superconducting state is fully characterized by its energy spectrum, as shown in Fig.~\ref{FIG1}. Therefore, we adopt the following thermodynamic formalism, which contains only the quasiparticle spectrum of the BdG Hamiltonian, to calculate the supercurrent. The supercurrent is defined as\cite{D-Xiao}
\begin{equation}
J_\alpha = \frac{\partial F}{\partial A^\alpha},
\label{thermo}
\end{equation}
with the free energy $F$ of the system given by
\begin{equation}
F= - k_B T \sum_{n{\bf k}}\ln[1+ e^{-({\tilde \epsilon}_n-\mu)/ k_B T}].
\label{free}
\end{equation}
Here, ${\tilde \epsilon}_n$ represents the energy spectrum modified by the external electric field, $A^{\alpha}$ denotes the $\alpha$ component of the vector potential, and $T$ is the temperature. $\alpha$, $\beta$ label Cartesian coordinates $x$, $y$, $z$. The vector potential ${A}^{\alpha} (t)$, which acts as the driving force, is related to the time-dependent electric field $E^\alpha(t)=E^\alpha e^{- i (\omega+i\eta)t}$ through $E^\alpha(t)=\partial A^\alpha (t)/\partial t$, where $\eta$ is a positive infinitesimal quantity to ensure the convergence at $t=-\infty$. Note that Eq.~(\ref{thermo}) is similar to the expression employed in investigating the superconducting diode effect.\cite{L-Fu2} From Eqs.~(\ref{thermo}) and (\ref{free}), the current density is expressed as
\begin{equation}
J_\alpha = \sum_{n{\bf k}} g_n \dfrac{\partial \tilde{\epsilon}_{n}}{\partial A^\alpha}, \label{current1}
\end{equation}
where $g_n$ is the nonequilibrium Fermi distribution. Note that Eq.~(\ref{thermo}) is based on a thermodynamic approach, which is suitable for equilibrium systems, consistent with the fact that the supercurrent is fundamentally an equilibrium property. However, in the gapless superconducting state, both quasiparticle current and supercurrent can coexist. To investigate the ScHE induced by a small electric field, where the system remains close to equilibrium, we continue to employ Eq.~(\ref{current1}), but replace the equilibrium Fermi distribution function $f_n$ with its nonequilibrium counterpart $g_n$.\cite{supple}


Eqs.~(\ref{thermo}) and (\ref{free}) are the standard thermodynamic formalism for calculating the supercurrent. In the following, we show that these equations can also reproduce the conventional current density expression for the normal current. We begin with the Hamiltonian in the velocity gauge,
\begin{equation}
H = H_{BdG} -\hat{v}^\alpha A^\alpha + \frac{1}{2} \hat{w}^{\alpha\beta} A^\alpha A^\beta. \label{ham1}
\end{equation}
Here $\hat{v}^\alpha = \partial H_{BdG}/\partial A^\alpha$ is the paramagnetic current operator while $\hat{w}^{\alpha\beta}=\partial^2 H_{BdG}/(\partial A^\alpha \partial A^\beta)$ is the diamagnetic current operator.~\cite{JEMooreSC,YanaseSC} To calculate current density in the linear response regime, it is sufficient to consider the second-order energy correction in Eq.~(\ref{ham1}), i.e., $\epsilon_n^{(2)}$. The derivation detail is provided in the Supplementary Material.~\cite{supple,J-Sipe}

In the dc limit ($\omega \rightarrow 0$) and under the relaxation time approximation, we finally obtain the current density in vector notation:
\begin{equation}
{\bf J} = \sum_{n{\bf k}} [ \tau {\bf E} \cdot \nabla_{\bf k} f_n {\boldsymbol v}_n - f_n {\bf E} \times {\bf \Omega}_n], \label{current3}
\end{equation}
where $\tau$ is the relaxation time and ${\bf \Omega}_n$ is the quasiparticle Berry curvature. The first term corresponds to the extrinsic Drude contribution. The second term is the intrinsic contribution from quasiparticle Berry curvature, inducing a dissipationless transverse supercurrent and hence the ScHE. It has been shown that quasiparticle Berry curvature plays a crucial role in the transport properties of superconductors.\cite{quasiparticleBC1,quasiparticleBC2,JEMooreSC,YanaseSC,CXiao2021} In Eq.~(\ref{current3}), the summation runs over all occupied bands and wave vectors ${\bf k}$ of the quasiparticles states. We emphasize that Eq.~(\ref{current3}) can be used to calculate both the quasiparticle current and the supercurrent for the gapless superconducting state. In the following, the second term of Eq.~(\ref{current3}) is adopted to study the intrinsic ScHE.

Figure~\ref{FIG1}(f) shows the Hall supercurrent density as a function of the orientation angle of the in-plane magnetic field for $E_F = 0$. When the magnetic field is along $y$ direction, i.e., $\theta = 0$, $J_H$ is zero. As the angle $\theta$ deviates from $0^\circ$, $J_H$ increases slowly. Moreover, $J_H$ increases with the warping strength $\lambda$, and vanishes at $\lambda = 0$. We confirm that quasiparticle Berry curvature also vanishes at $\lambda = 0$, providing strong evidence that the transverse supercurrent $J_H$ arises from the quasiparticle Berry curvature effect. As presented in Fig.~\ref{FIG1}(b), when the orientation angle $|\theta|$ is larger than $30^\circ$ at $E=0$, the superconducting gap is closed, suppressing the formation of supercurrent and resulting in normal current of quasiparticles. As shown below, the results shown in Fig.~\ref{FIG1}(f) can also be consistently reproduced using quantum transport theory.

%

\begin{figure}
\centering
\includegraphics[width=\columnwidth]{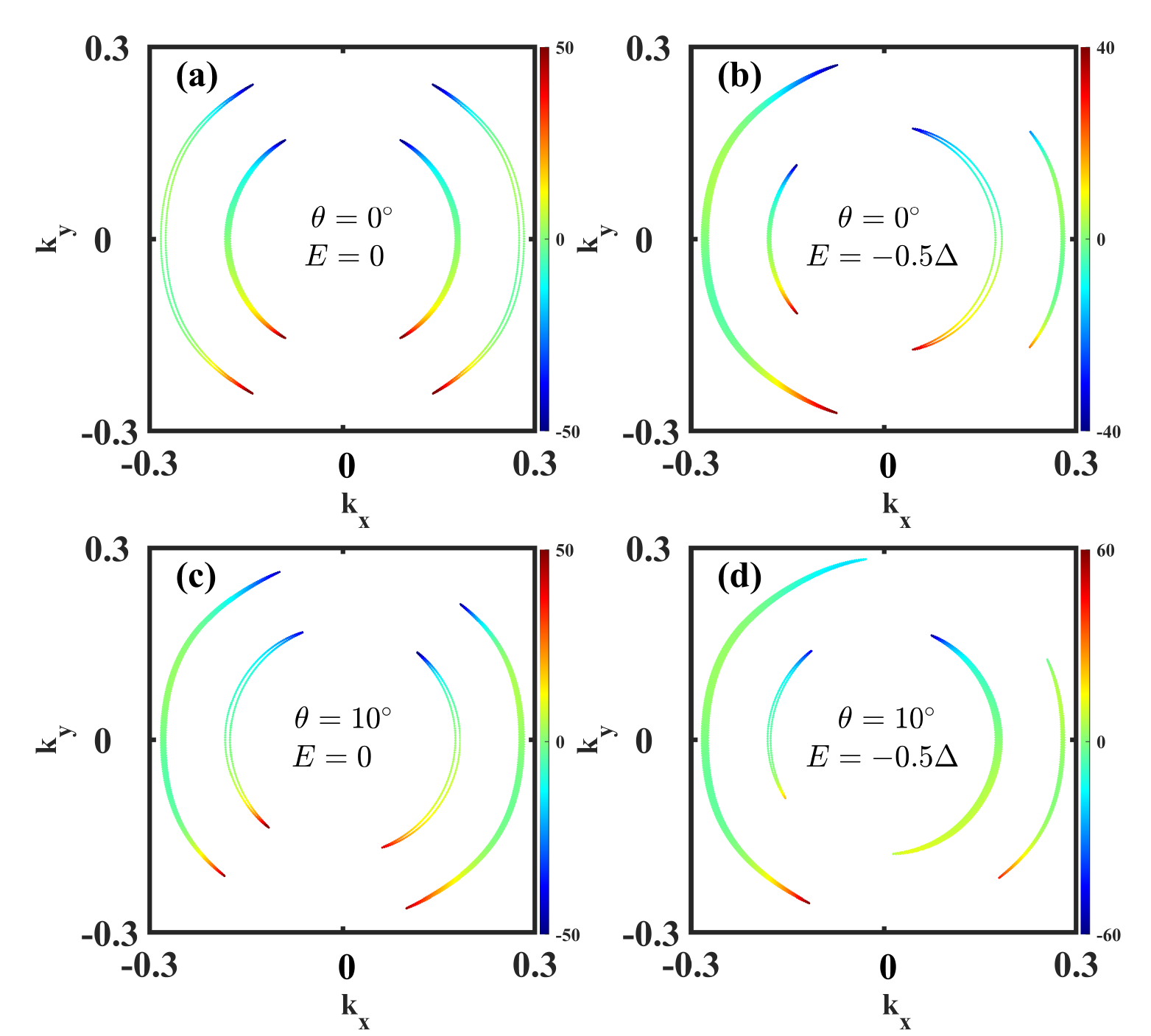}
\caption{ Quasiparticle Berry curvature distribution on the segmented Fermi surfaces for different orientation angles of the magnetic field and energies: (a) $\theta=0^\circ$ and $E=0$; (b) $\theta=0^\circ$ and $E=-0.5\Delta$; (c) $\theta=10^\circ$ and $E=0$; (d) $\theta=10^\circ$ and $E=-0.5\Delta$. Quasiparticle Berry curvature is absent within the superconducting gap. Other parameters are the same as Fig.~\ref{FIG1}.}
\label{FIG2}
\end{figure}

To further clarify the physical origin of the ScHE, Fig.~\ref{FIG2} shows the distribution of quasiparticle Berry curvature for different energy $E$ and angle $\theta$. At $\theta = 0^\circ$, the Berry curvature distribution is symmetric with respect to $k_x$ and antisymmetric with respect to $k_y$ at $E = 0$ [Fig.~\ref{FIG2}(a)].\cite{supple} For energies below $E=0$, the symmetry in $k_x$ is broken, but the antisymmetry in $k_y$ persists [Fig.~\ref{FIG2}(b)]. Consequently, the transverse supercurrent vanishes at $\theta = 0^\circ$ due to the antisymmetric nature of the Berry curvature where $\Omega(k_y) = -\Omega(-k_y)$. At $\theta = 10^\circ$, the Berry curvature is asymmetric with a symmetry axis rotated by $10^\circ$ at $E=0$ [Fig.~\ref{FIG2}(c)]. This is because $E=0$ is a special point that preserves electron-hole symmetry. However, below $E=0$, the Berry curvature generally exhibits an asymmetric pattern, as shown in Fig.~\ref{FIG2}(d), resulting in a nonzero transverse supercurrent.

\begin{figure}[t]
\centering
\includegraphics[width=0.90\columnwidth]{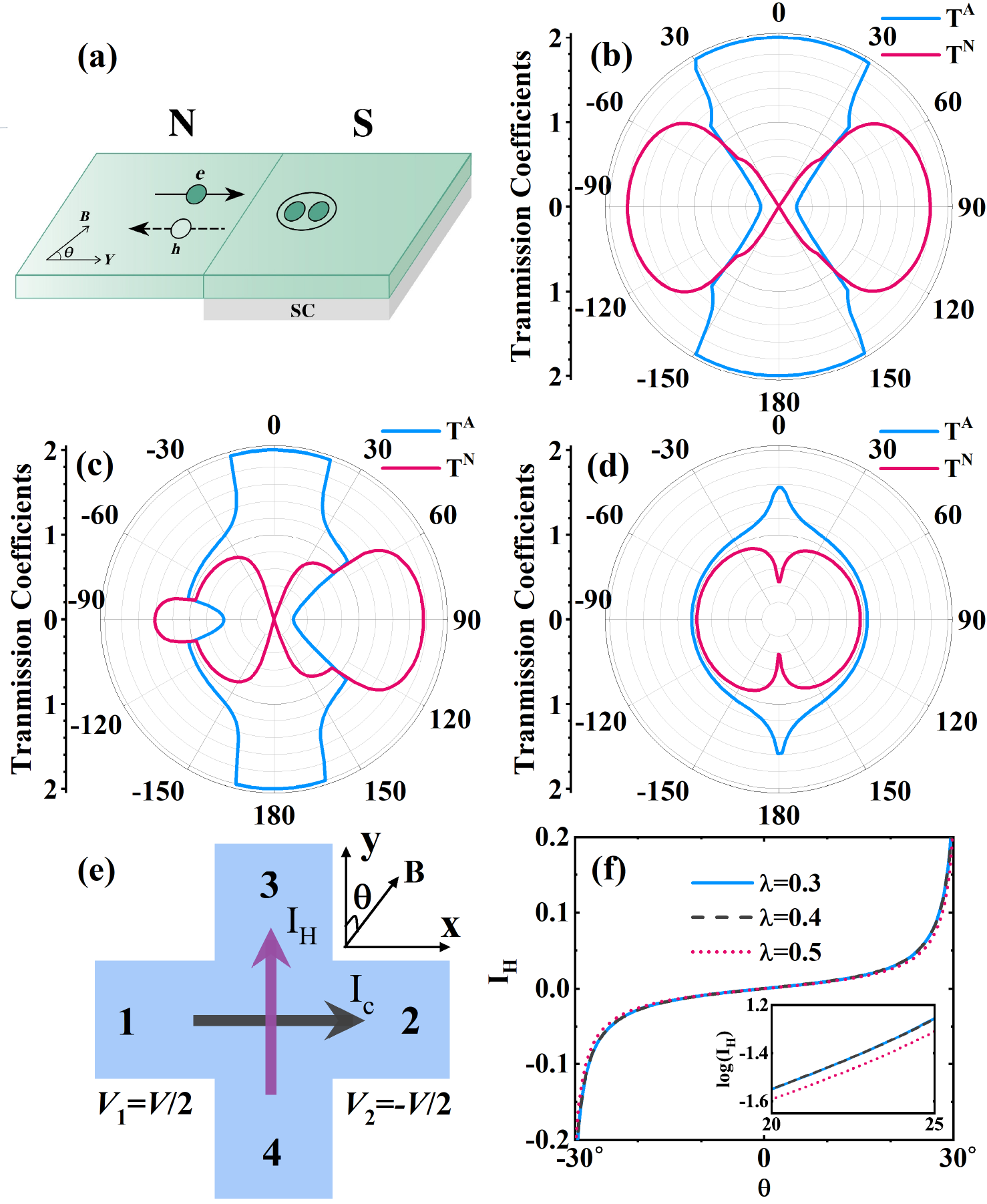}
\caption{(a) Schematic of a two-probe N-S hybrid system. (b)-(d) Angle-resolved transmissions $T_N$ and $T_A$ at different energies $E = 0$, $-0.5\Delta$, and $-\Delta$, respectively. (e) Schematic of a four-probe setup based on the gapless superconductor, where the quasiparticle current $I_c$ flows along the $x$ direction and the supercurrent flows along the $y$ axis. (f) Transverse suppercurrent $I_H$ as a function of $\theta$ with various $\lambda$ in the four-probe setup. Other parameters are the same as Fig.~\ref{FIG1}.}\label{FIG3}
\end{figure}

\bigskip
\noindent{\it Quantum transport formalism} --- We now investigate the ScHE from the perspective of quantum transport. We begin by analyzing quantum transport in a two-probe normal metal-superconductor (N-S) system, as illustrated in Fig.~\ref{FIG3}(a), to show that the transport calculation gives the same information on segmented Fermi surface in Fig.~\ref{FIG1}. The superconductor region is governed by the Hamiltonian in Eq.~(\ref{ham}), while the normal metal region follows the same Hamiltonian with $\Delta=0$. As the in-plane magnetic field rotates, the superconducting gap associated with the segmented Fermi surface rotates accordingly in the $k_x$-$k_y$ plane. At the N-S interface, both normal transmission and Andreev reflection may occur, depending on the orientation of the superconducting gap. When the gap opens along the $k_y$ direction, Andreev reflection dominates; otherwise, quasiparticle transmission is the primary transport mechanism. The normal transmission coefficient $T^N$ and Andreev reflection coefficient $T^A$ are given by\cite{Xing,Sun1}
\begin{eqnarray}
T^N &=& \textbf{Tr} \left[ \Gamma_{Le} (G^r \Gamma_R G^a)_{ee} \right] \nonumber \\
T^A &=& \textbf{Tr} \left[ \Gamma_{Le} G^r_{eh} \Gamma_{Lh} G^a_{he} \right],
\label{trans}
\end{eqnarray}
where $G^r = [G^r_{ee}, G^r_{eh}; G^r_{he}, G^r_{hh}]$ is the retarded Green's function of the N-S system. The subscript $e$ ($h$) denotes the electron (hole) component. $\Gamma_{Le}(E)=\Gamma_{Lh}(-E)$ represents the electron and hole linewidth functions in the left normal-metal lead, and $\Gamma_{R}$ denotes the linewidth function of the right superconducting lead.\cite{Xing,Sun1}

Figure~\ref{FIG3}(b)-(d) illustrate the angle-resolved $T^N$ and $T^A$ at different energies. Two observations are in order: (1) As far as the superconducting gap is concerned, Fig.~\ref{FIG1} and Fig.~\ref{FIG3} agree with each other very well; (2) For the chosen parameters, two transmission channels are available, yielding $T^N + T^A = 2$, which ensures current conservation. In particular, when the superconducting gap opens along $k_y$ direction ($\theta=0$), transport occurs purely via Andreev reflection with $T^A = 2$, while normal transmission is forbidden with $T^N=0$. The transmission coefficients exhibit excellent agreement with the features of segmented Fermi surfaces depicted in Fig.~\ref{FIG1}(b)-(d).

We further calculate the transverse supercurrent for a four-probe setup, as illustrated in Fig.~\ref{FIG3}(e). The system is governed by the gapless superconductor Hamiltonian in Eq.~(\ref{ham}), with the magnetic field oriented along the \( y \) axis. In this case, when a longitudinal quasiparticle current is driven by the bias voltages in the \(x\) direction, tranverse supercurrent flows along the \( y \) direction, similar to the scenario shown in Fig.~\ref{FIG1}(e). For a multi-probe system, the current flowing through lead $\alpha$ can be calculated as\cite{DattaBook}
\begin{equation}
I_{\alpha}=-2\int \frac{dE}{2\pi}{\bf Tr}\left[ {{G}^{r}}\Sigma_{\alpha}^{<}-\Sigma_{\alpha}^{<}{{G}^{a}}+{{G}^{<}}\Sigma_{\alpha}^{a}-\Sigma_{\alpha}^{r}{{G}^{<}} \right]_{ee},
\label{Ialpha}
\end{equation}
where the Green's functions (${G}^{r}$, $G^{a}$, ${{G}^{<}}$) and the self-energies ($\Sigma_{\alpha}^{r}$, $\Sigma_{\alpha}^{<}$) are defined in a standard way.\cite{DattaBook} The factor $2$ accounts for contributions from both electron and hole. Similar to the thermodynamic approach in Eq.~(\ref{current3}), Eq.~(\ref{Ialpha}) can also calculate both normal current and supercurrent.


Figure~\ref{FIG3}(f) presents the dependence of the transverse supercurrent $I_H$ (equal to $I_3$ or $I_4$) on the magnetic field orientation angle $\theta$ and the warping strength $\lambda$. Two prominent features are observed: (1) $I_H$ vanishes at $\theta = 0^\circ$ and increases monotonically as $\theta$ deviates from zero; (2) $I_H$ is zero when $\lambda = 0$, but becomes finite for $\lambda \neq 0$. These characteristics of $I_H$ are qualitatively consistent with those of the current density $J_H$ shown in Fig.~\ref{FIG1}(f), supporting the existence of the ScHE. Quantitative differences between $I_H$ and $J_H$ are expected, as the former represents the total current, whereas the latter denotes the current density.



\smallskip
\noindent{\it Experimental realization}---
Experimental verification of the intrinsic ScHE would be straightforward. The gapless superconducting state has been realized in Bi$_2$Te$_3$/NbSe$_2$ heterostructures,\cite{J-Jia} where several layers of Bi$_2$Te$_3$ are epitaxially grown on the superconducting substrate NbSe$_2$. To detect the ScHE, one can fabricate a circular disc setup by attaching multiple probing electrodes (twelve or more) to the Bi$_2$Te$_3$/NbSe$_2$ sample, which has been a well-established technique in Hall transport measurements.\cite{KKang2019,Kumar2021,Gao2021} By performing the angular-dependent Hall measurement, in which probing electrodes are rotated around the disc, the angle-resolved transport response can reveal signatures of both the tilted superconducting gap [Fig.~\ref{FIG3}(b)-(d)] and the ScHE of the gapless superconducting state. Alternatively, rotating the in-plane magnetic field provides another means of manipulating the transverse supercurrent. A schematic of the circular disc setup and the rotating-electrode measurement is shown in the Supplementary Material.\cite{supple} It is worth noticing that there are no voltage drop across the superconducting electrodes, which can instead be detected via Andreev reflection signals.\cite{Du2012}


\smallskip
\noindent{\it Conclusion} --- In summary, we propose an unconventional Hall effect in the gapless superconducting state with segmented Fermi surfaces, which is referred to as the ScHE. In this effect, a longitudinal quasiparticle current is converted into a transverse supercurrent carried by finite-momentum Cooper pairs. To elucidate the underlying mechanism, we develop a thermodynamic approach that treats quasiparticle current and supercurrent on equal footing, and identity the essential role of quasiparticle Berry curvatures in generating the intrinsic ScHE. Quantum transport calculations for a four-probe setup further confirm the existence of this ScHE. Finally, we outline experimental strategies for detecting the ScHE in proximitized gapless superconductors.

\smallskip
\noindent{\it Acknowledgement}--- This work is supported by the National Natural Science Foundation of China (Grants No. 12034014, No. 12404058, No. 12404059, and No. 12174262). B.W. thanks the Shenzhen Natural Science Foundation (Grant No. 20231120172734001).

\end{document}